\documentclass[aps,pre,twocolumn,amsmath,superscriptaddress,showpacs]{revtex4}

\usepackage{amssymb}

\allowdisplaybreaks[1]

\begin{document}

\title{Temporal Correlations of Local Network Losses}

\author{A. S. Stepanenko}
\affiliation{School of Engineering, University of Birmingham, Edgbaston, Birmingham, B15 2TT, UK}

\author{C. C. Constantinou}
\affiliation{School of Engineering, University of Birmingham, Edgbaston, Birmingham, B15 2TT, UK}

\author{I. V. Yurkevich}
\affiliation{School of Physics and Astronomy, University of Birmingham, Edgbaston, Birmingham, B15 2TT, UK}

\author{I. V. Lerner}
\affiliation{School of Physics and Astronomy, University of Birmingham, Edgbaston, Birmingham, B15 2TT, UK}

\date{\today}

\begin{abstract}
We introduce a continuum model describing data losses in a single node  of a packet-switched network (like the Internet) which preserves the discrete nature of the data loss process. {\em By construction}, the model has critical behavior with a sharp transition from exponentially small to finite losses with increasing data arrival rate.  We show that such a model  exhibits strong fluctuations in the loss rate at the critical point  and non-Markovian power-law correlations in time, in spite of the Markovian character of the data arrival process. The continuum model allows for rather general incoming data packet distributions and can be naturally generalized to consider the buffer server idleness statistics.
\end{abstract}


\maketitle

\section{Introduction}

Complex networks underpin many diverse areas of science. They manifest themselves in relationships between network topology and functional organization of complex neuron structures \cite{Zhou:06,Arbib:01}, interacting organic molecules describing metabolic activity in living cells \cite{Jeong:00}, multi-species food webs \cite{Kim:05,Cohen:90}, numerous aspects of social networks \cite{Li:07,Lind:07,Gallos:07,Liljeros:01}, and the connectivity and operation of the Internet \cite{Kurant:07,Duch:06,Pastor:01}. New models of network topology such as scale-free \cite{Barabasi:99} or small-world \cite{Watts:98} have been found to be surprisingly good at describing real-world structures. A consequence of the realisation that complex networks describe universal properties of many such  problems has resulted in extensive research activity by the physics community in the past decade (see Refs.~\cite{Albert:02,Watts:99} for reviews).

A problem of particular significance in many application domains is the resiliency of complex networks to the random or selective removal of nodes or links. For example, the loss of connectivity in scale-free networks \cite{Kurant:07,Braunstein:03,Albert:00,Cohen:00,Dorogovtsev:00} has implications on the tolerance of the Internet to protocol or equipment failures. Typically, the site or bond disorder acts as an \textit{input} which makes them very general and applicable to a wide variety of networks.

More recently there has been an increasing realization that network breakdowns can not only result from the physical loss of connectivity, but can arise due to the loss of data traffic in the network (i.e. congestion) \cite{Ashton:05,Guimera:02}. However, only a few dynamical models of traffic in networks have been considered to date \cite{Duch:06,Stepanenko:05,Menezes:04,Stepanenko:02}. In the case of communication networks the excessive loading of even a single node can give rise to cascades of failures arising from traffic congestion and consequently isolate large parts of the network \cite{Moreno:03}. To describe the operational failure arising due to congestion at a particular network node, one needs to account for distinct features of the dynamically `random' data traffic which is the reason for such a breakdown.

In this paper we model data losses in a \textit{single node} of a packet-switched network like the Internet. There are two distinct features which must be preserved in this case: the discrete character of data propagation and the possibility of data overflow in a single node.  In {the packet-switched network} data is divided into packets which are routed from source to destination via a set of interconnected nodes (routers). At each node packets are queued in a memory buffer before being {\it serviced}, i.e.\ forwarded to the next node (there are separate buffers for incoming and outgoing packets but we neglect this for the sake of simplicity). Due to the finite capacity of memory buffers and the stochastic nature of data traffic, any buffer can become overflown which results in  packets being {\it discarded}.

We focus on a continuum description of the discrete process of data packet loss. Such a continuum model represents a simplification that preserves the salient features of the data loss mechanism, while at the same time it can be more easily embedded in a larger model describing data packet losses in a large network. The continuum description allows us to overcome inevitable difficulties in incorporating realistic distributions of incoming traffic into a discrete-time class of models, like one we introduced earlier  \cite{Yur06}. On the contrary, the continuum model can  easily incorporate a  completely general distribution of packet lengths and inter-arrival times, both essential in modeling data loss in finite-sized buffers.

We introduce a model where noticeable data losses in a single memory buffer start when the average rate of random packet arrivals approaches the service rate. Under this condition the model has {\em a built-in sharp} transition from free flow to lossy behavior with a sizeable fraction of arriving packets being dropped. A sharp onset of network congestion is familiar to everyone using the Internet and was numerically confirmed in different models \cite{Ohira:98}. Here we stress that such a congestion originating from a single node is characterized by strong critical fluctuations of the data loss in the vicinity of the built-in transition.

In particular, we will show that a Markovian input process can give rise to long-range temporal correlations of data losses that are strongly non-Markovian in the critical regime.  In the context of the Internet, this means that when excessive data losses start  it is more probable that they persist for a while, thus impacting on network operation.  As we will discuss later in this paper, this non-Markovian behavior has a profound effect on the operation of current Internet protocols, such as the Transport Control Protocol (TCP), that dictate how users experience the network operation.

While data loss is natural and inevitable due to data overflow, we show that loss rate statistics  turn out to be highly nontrivial in the realistic case of a finite buffer, where at the critical point  the magnitude of fluctuations can exceed the average value. The fluctuations still obey the central limit theorem but  only in the unrealistically long time limit. The importance of fluctuations in some intermediate regime is a definitive feature of {\it mesoscopic} physics, albeit the reasons for this are absolutely different (note that even in the case of electrons, the origin of the mesoscopic phenomena can be either quantum or purely classical, see, e.g., \cite{IVL:93}).

The {\it average} loss rate and/or transport delays were previously studied, e.g.,  in the {theories} of bulk queues \cite{Cohen:69,Schwartz:87} {or a jamming transition in traffic flow} \cite{Cates:98}. What makes present considerations intrinsically  different from these theories is the very nature of the quantity we consider: the losses (not existing in flow models) make the description of the traffic process essentially non-Hermitian.
Although fluctuations in network dynamics were previously studied  (see, e.g.\ \cite{Duch:06,Menezes}), this was done  through measurements or numerical simulations of   data traffic.

Due to the symmetry of the continuum description of a buffer with respect to its full (lossy) and empty (idle) states, we also derive corresponding expressions for the statistics of idleness of the buffer server (i.e.\ output links from routers). This quantity is essential in determining the way the statistics of data traffic going into a subsequent buffer along a data path are shaped. This is self-evidently important when we are attempting to describe the operation of an entire network.

\section{The Model}

We consider a single finite-size memory buffer fed with a random data-packet stream. It stores the packets and then is \textit{serviced} by the data-link that sends this packets further along the network on a first-in-first-out  basis. This adequately models the output buffer attached to the switching device in the router. The speed of the input line of the buffer is   much bigger than the speed of the output line. The reason is that the input comes from the switching fabric of a router which is designed to operate very fast indeed in order to feed a large number of such buffers, but sequentially. The capacity of the output line is normally smaller.

 Hence, we can model the packet arrival as an instantaneous renewal  process. The storage capacity of the buffer is $L$, measured in bits. The lengths of arriving packets are treated as random, all being much smaller than $L$. The service rate (i.e.\ the rate at which packets depart from the buffer) is considered to be deterministic, as randomness in it is negligible as compared to that of the input traffic. We normalize the lengths of packets $p$, the speed of the output link $r_{\rm out}$ and the queue length $\ell$ by the size of the buffer $L$ (which  is henceforth set to $1$).

The procedure for the renewal cycle  is described as follows: at the moment of arrival of a packet of size $p$, the state of the queue is
$\ell$, this is followed by the gap $\eta$ (random inter-arrival time) until the next arrival. We introduce  the  time scale required to empty a full buffer provided there are no new arrivals,
$
  \eta_0\equiv {1}/{r_{\rm out}}.
$ If $\ell+p\le 1$ then the
packet joins the queue and the queue length prior the next arrival is $\ell'=\ell+p-\eta/\eta_0$ if $\ell'>0$ and
$\ell'=0$ otherwise. If $\ell+p > 1$ then the packet is discarded and the queue length prior the next arrival is
$\ell'=\ell-\eta/\eta_0$ if $\ell'>0$ and $\ell'=0$ otherwise.

Since the maximum packet size is much less than $1$ (the buffer size) and assuming that the average incoming traffic rate
$r_{\rm in}$ (also normalized to the buffer size) is close to the service rate:
\begin{equation}
  | r_{\rm in}\eta_0 - 1 | \ll 1
\label{eq0}
\end{equation}
we can treat $p$, $\eta$ and $\ell$ as continuous variables.

Our aim is to calculate the statistics of the amount of the dropped traffic and the service lost due to idleness of the
output link during time $t\gg\bar\eta$ ($\bar\eta$ is the average inter-arrival time) in the regime (\ref{eq0}).
In this regime  and for observation times $t\gg\bar\eta$, the system can be described by the Fokker-Planck
equation as follows (in terms of the transitional probability density function $w(\ell',t;\ell)$)
\begin{equation}
 \partial_t w(\ell',t;\ell)
  = - a \partial_{\ell'} w(\ell',t;\ell)
    + \frac{1}{2}\sigma^2\partial_{\ell'}^2 w(\ell',t;\ell)\ ,
\label{FP}
\end{equation}
where $a$ and $\sigma^2$ are average moments of the change of the queue size per unit time
\begin{equation}
 a \equiv\frac{1}{\Delta t}\langle \Delta \ell \rangle\ , \quad
 \sigma^2 \equiv\frac{1}{\Delta t}\langle \Delta \ell^2 \rangle\ ,\quad \Delta t\to 0
\label{moments}
\end{equation}
and the following boundary and initial conditions are imposed
\begin{equation}
 \left. J(\ell',t;\ell)\right|_{\ell'=0,1} = 0\ , \quad
\label{boundary}
\end{equation}
\begin{equation}
 \left.w(\ell',t;\ell)\right|_{t=0} = \delta(\ell'-\ell)\ ,
\label{initial}
\end{equation}
where
\begin{equation}
 J(\ell',t;\ell)\equiv a w(\ell',t;\ell) - \frac{1}{2}\sigma^2\partial_{\ell'} w(\ell',t;\ell)
\label{current}
\end{equation}
is the probability current. By $\Delta t\to 0$ in eq.~(\ref{moments}) we mean that $\Delta t$ is much smaller than the
observation time, but large enough so that the underlying stochastic processes can be considered as continuous:
\begin{equation}
 \bar\eta\ll\Delta t\ll t
\label{time}
\end{equation}

The solution of (\ref{FP},\ref{boundary},\ref{initial}) can be expressed as follows
\begin{equation}
\begin{split}
 w(\ell',t;\ell)
  =& 2{\rm e}^{v(\ell'-\ell)}\sum\limits_{k=1}^\infty
  \frac{\exp\left[-(4\pi^2k^2 + v^2)\tau\right]}{4\pi^2k^2 + v^2}
\\
&\quad\times
  \left[ 2\pi k\cos(2\pi k\ell') + v \sin(2\pi k\ell') \right]
\\
&\quad\times
  \left[ 2\pi k\cos(2\pi k\ell) + v \sin(2\pi k\ell) \right]
\label{eq1a}
\end{split}
\end{equation}
where
\begin{equation}
 v\equiv\frac{a}{\sigma^2}\ ,\qquad
 \tau\equiv\frac{\sigma^2t}{2}
\label{eq1b}
\end{equation}
Note that the solution (\ref{eq1a}) can be expressed in terms of $\theta$-functions.

For the Laplace transform of $w(\ell',t;\ell)$ we have
\begin{equation}
\begin{split}
 &W(\ell',\epsilon;\ell)
  \equiv {\cal L}_\tau w(\ell',t;\ell)
 = \frac{1}{2}\frac{{\rm e}^{v(\ell'-\ell)}}{\kappa\sinh(\kappa)}
\\
  &\quad \times
  \Biggl\{
     \frac{2v^2}{\epsilon}\cosh[\kappa(\ell'+\ell-1)]
   + \frac{2\kappa v}{\epsilon}\sinh[\kappa(\ell'+\ell-1)]
\\
&\qquad
   + \cosh[\kappa(|\ell'-\ell|-1)] + \cosh[\kappa(\ell'+\ell-1)]
  \Biggr\}
\label{eq2}
\end{split}
\end{equation}
where
\begin{equation}
 \kappa\equiv\sqrt{\epsilon + v^2}
\label{eq3}
\end{equation}
From (\ref{eq2}) we have for the probabilities of returning to the boundaries
\begin{equation}
\begin{split}
 W(0,\epsilon;0) &= \frac{1}{\epsilon}\left[ \kappa\,{\rm cotanh}(\kappa) - v \right]
\\
 W(1,\epsilon;1) &= \frac{1}{\epsilon}\left[ \kappa\,{\rm cotanh}(\kappa) + v \right]
\end{split}
\label{eq4}
\end{equation}
These will be used in the next section.

\section{Statistics of losses}

In this section we concentrate on the statistics of the losses due to the buffer overflowing. The corresponding
formulae for the statistics of the server idleness can be obtained using transformation $\ell\to1-\ell,v\to-v$.

First, we estimate the size of fluctuations of the losses on a time scale $t\ll 2/\sigma^2$. In order to do that we
consider the dynamics of the system near the boundary $\ell=1$ which is governed by the following transitional
probability:
\begin{equation}
\begin{split}
 & w_0(\ell',t;\ell)
 = \frac{1}{\sqrt{2\pi\sigma^2 t}}\exp\left[ - \frac{a(\ell'-\ell)}{\sigma^2} - \frac{a^2t}{2\sigma^2} \right]
\\
&\quad \times
    \left\{
       \exp\left[ - \frac{(\ell'-\ell)^2}{2\sigma^2 t} \right]
     + \exp\left[ - \frac{(2-\ell'-\ell)^2}{2\sigma^2 t} \right]
    \right\}
\\
&\quad
 - \frac{a}{\sigma^2}\exp\left[ \frac{2a(1-\ell')}{\sigma^2} \right]
    {\rm erfc}\left[ \frac{2-\ell'-\ell+at}{\sqrt{2\sigma^2t}} \right]
\end{split}
\label{eq7a}
\end{equation}
which is the solution of (\ref{FP}) when the boundary $\ell=0$ is sent to $-\infty$. The change in the state of the
system during time $t$ can then be represented as follows:
\begin{equation}
 \Delta\ell(t)\equiv \ell' - \ell = \Delta\ell_0(t) + \Delta\ell_{\rm loss}(\ell',t;\ell)
\label{eq6}
\end{equation}
where $\Delta\ell_0(t)$ is the change in the state of the system if there was no boundary, its statistics is determined
by
\begin{equation}
 \langle\Delta\ell_0(t)\rangle = at\ ,\quad
 \langle[\Delta\ell_0(t)]^2\rangle = \sigma^2t + {\rm o}(t)\ ,\quad
\label{eq7}
\end{equation}
and $\Delta\ell_{\rm loss}(\ell',t;\ell)$ is the amount of traffic lost due to buffer overflowing. The moments of
(\ref{eq7}) can be defined as follows
\begin{equation}
 \langle[\Delta\ell(t)]^n\rangle
  = \int\!{\rm d}\ell'{\rm d}\ell\ (\ell'-\ell)^nw_0(\ell',t;\ell)p(\ell)
\label{eq7b}
\end{equation}
where $p(\ell)$ is the stationary distribution of buffer occupancy.

For the first two moments (\ref{eq7b}) in the limit $t\to0$ we have
\begin{equation}
 \langle\Delta\ell(t)\rangle = at + \frac{\sigma^2t}{2}p(1)\ ,\quad
 \langle[\Delta\ell(t)]^2\rangle = \sigma^2t
\label{eq7c}
\end{equation}
From (\ref{eq6},\ref{eq7},\ref{eq7c}) we can conclude that
\begin{equation}
\begin{split}
 &\langle\Delta\ell_{\rm loss}(t)\rangle = \frac{\sigma^2t}{2}p(1)
\\
 &\langle[\Delta\ell_{\rm loss}(t)]^2\rangle + 2\langle\Delta\ell_0(t)\Delta\ell_{\rm loss}(t)\rangle = {\rm o}(t)
\end{split}
\label{eq7d}
\end{equation}
The first of the relations (\ref{eq7d}) means that $\Delta\ell_{\rm loss}(\ell',t;\ell)$ is non-zero only if
$\ell',\ell\sim1$ in the limit $t\to0$. The second relation means either
\begin{equation}
 \langle[\Delta\ell_{\rm loss}(t)]^2\rangle ,\langle\Delta\ell_0(t)\Delta\ell_{\rm loss}(t)\rangle = {\rm o}(t)
\label{eq7e}
\end{equation}
or
\begin{equation}
 \Delta\ell_{\rm loss}(t) = - 2\Delta\ell_0(t) + {\rm o}(\sqrt{t})
\label{eq7f}
\end{equation}
The relation (\ref{eq7f}) does not make sense physically, so in what follows we accept option (\ref{eq7e}) and show
that it is consistent with the later calculations.

Next we lift the restriction $t\ll 2/\sigma^2$. It can be shown that the conditional moments (with the condition that
the system was in the state $\ell$ at the beginning of the observation interval) can be expressed as follows:
\begin{equation}
\begin{split}
 & m_{\rm loss}^{(k)}(t;\ell)
 = k!r_{\rm loss}^k \prod_{i=1}^{k} \int\limits_{0}^{t_{i+1}}\!{\rm d}t_i
    \prod_{j=1}^{k-1} w(1,t_{j+1} - t_j;1)
\\
&\qquad\qquad\quad\times
    w(1,t_1;\ell)\ ,\quad t_{k+1}\equiv t
\end{split}
\label{eq11}
\end{equation}
where $w(\ell',t;\ell)$ is determined by (\ref{eq1a}) and
\begin{equation}
\begin{split}
 r_{\rm loss}
  &\equiv
    \lim_{t\to 0}\frac{1}{t}\int\!{\rm d}\ell'\int\!{\rm d}\ell\ \Delta\ell_{\rm loss}(\ell',t;\ell)
\\
  &=
    \lim_{t\to 0}\frac{1}{t}\int\limits_{-\infty}^1\!{\rm d}\ell'{\rm d}\ell\ (\ell'-\ell-at)w_0(\ell',t;\ell)
  = \frac{\sigma^2}{2}
\end{split}
\label{eq12}
\end{equation}
For unconditional moments in the stationary regime we have
\begin{equation}
\begin{split}
 m_{\rm loss}^{(k)}(t)
  &\equiv \int\limits_0^1\!{\rm d}\ell\ m_{\rm loss}^{(k)}(t;\ell) p(\ell)
\\
  &= k! \prod_{i=1}^{k} \int\limits_{0}^{\tau_{i+1}}\!{\rm d}\tau_i
    \prod_{j=1}^{k-1} w(1,t_{j+1} - t_j;1)\cdot p(1)
\end{split}
\label{eq13}
\end{equation}
where $\tau$ is defined in (\ref{eq1b}) and $p(\ell)$ is the stationary solution of (\ref{FP}):
\begin{equation}
 p(\ell) = \frac{2v{\rm e}^{2v\ell}}{{\rm e}^{2v}-1}
\label{eq14}
\end{equation}

To calculate $m_{\rm loss}^{(k)}(t)$ we consider its Laplace transform:
\begin{equation}
\begin{split}
 M_{\rm loss}^{(k)}(\epsilon)
  &\equiv {\cal L}_\tau m_{\rm loss}^{(k)}(t)
   = \int\limits_0^\infty{\rm d}\tau\ {\rm e}^{-\epsilon\tau} m_{\rm loss}^{(k)}(t)
\\
 &= k! p(1)\left[ {\cal L}_\tau w(1,t;1) \right]^{k-1} {\cal L}_\tau \tau
\\
 &= k! p(1) \left[W(1,\epsilon;1)\right]^{k-1} \frac{1}{\epsilon^2}
\end{split}
\label{eq16}
\end{equation}
where $W(1,\epsilon;1)$ is is defined by (\ref{eq2}).

Taking now the inverse Laplace transform we have
\begin{equation}
 m_{\rm loss}^{(k)}(t)
  \equiv {\cal L}_\epsilon^{-1} M_{\rm loss}^{(k)}(\epsilon)
   = \frac{1}{2\pi{\rm i}}\int\limits_{\gamma-{\rm i}\infty}^{\gamma+{\rm i}\infty}{\rm d}\epsilon\
     {\rm e}^{\epsilon \tau} M_{\rm loss}^{(k)}(\epsilon)
\label{eq18}
\end{equation}
From (\ref{eq16}) we obtain
\begin{equation}
 m_{\rm loss}^{(1)}(t) = p(1) \tau = p(1) \frac{\sigma^2 t}{2}
\label{eq19}
\end{equation}
For the moments (\ref{eq16}) with $k>1$ we can identify the following regimes:
\begin{equation}
 M_{\rm loss}^{(k)}(\epsilon)
  =
  \begin{cases}
   k! p(1) \epsilon^{-(k+3)/2} & \epsilon\gg1 \\
   k! p^k(1) \epsilon^{-(k+1)}  & \epsilon\ll 1
  \end{cases}
\label{eq20}
\end{equation}
Correspondingly, for the moments in $t$-representation we have
\begin{equation}
 m_{\rm loss}^{(k)}(t)
  =
  \begin{cases}
   k! p(1) \displaystyle\frac{\tau^{(k+1)/2}}{\Gamma[(k+3)/2]} & \tau\ll1 \\
   p^k(1) \tau^k & \tau\gg 1
  \end{cases}
\label{eq21}
\end{equation}

Now we calculate the PDF $p_{\rm loss}(x;t)$ of the amount of the lost traffic, $x$,
during time $t$. To calculate it we consider its characteristic function in the $\epsilon$-representation:
\begin{equation}
\begin{split}
&
 \tilde P_{{\rm loss}}(s;\epsilon)
  \equiv {\cal L}_x P_{\rm loss}(x;\epsilon)\ ,\
\\
&
 P_{{\rm loss}}(x;\epsilon)
  \equiv {\cal L}_\tau p_{\rm loss}(x;t)
\end{split}
\label{eq31}
\end{equation}
From (\ref{eq31}) we obtain
\begin{equation}
\begin{split}
 \tilde P_{{\rm loss}}(s;\epsilon)
  &= \sum\limits_{k=0}^\infty\frac{(-s)^k}{k!}\int\limits_0^\infty\!{\rm d}x\ x^k {\cal L}_\tau p_{\rm loss}(x;t)
\\
  &= P_{\rm loss}(\epsilon) + \sum\limits_{k=1}^\infty\frac{(-s)^k}{k!} M^{(k)}_{\rm loss}(\epsilon)
\end{split}
\label{eq32}
\end{equation}
where
\begin{equation}
 P_{\rm loss}(\epsilon) = {\cal L}_\tau p_{\rm loss}(t)\ ,\ \
 p_{\rm loss}(t) = \int\limits_0^\infty\!{\rm d}x\ p_{\rm loss}(x,t)
\label{eq32a}
\end{equation}
with $1-p_{\rm loss}(t)$ being the probability for the system not to drop a single packet over the period of time $t$.
Substituting (\ref{eq16}) into (\ref{eq32}) we have
\begin{equation*}
\begin{split}
&
 \tilde P_{{\rm loss}}(s;\epsilon)
  = P_{\rm loss}(\epsilon) + \frac{p(1)}{\epsilon^2}\sum\limits_{k=1}^\infty (-s)^k [W(1,\epsilon;1)]^{k-1}
\\
&\
  = P_{\rm loss}(\epsilon) + \frac{p(1)}{\epsilon^2W(1,\epsilon;1)}
     \left[ - 1 + \frac{1}{1+sW(1,\epsilon;1)} \right]
\end{split}
\end{equation*}
In order that $P_{{\rm loss}}(s;\epsilon)$ did not have an abnormal behaviour (in particular, it did not contain terms
like $\delta(x)$), we must assume that
\begin{equation}
 P_{\rm loss}(\epsilon) = \frac{p(1)}{\epsilon^2W(1,\epsilon;1)}
\label{eq34}
\end{equation}
Hence,
\begin{equation}
 P_{{\rm loss}}(x;\epsilon) = \frac{p(1)}{\epsilon^2W^2(1,\epsilon;1)}
     \exp\left[ \frac{x}{W(1,\epsilon;1)} \right]
\label{eq35}
\end{equation}
Integrating this relation over $x$, we recover (\ref{eq34}), which shows that our assumption is indeed correct.

In the regimes of short and long times we have
\begin{equation}
 p_{{\rm loss}}(x;t)
  =
  \begin{cases}
   p(1) {\rm erfc}\left[\displaystyle\frac{x}{\sqrt{4\tau}}\right] & \tau\ll1 \\[4mm]
   \delta\Bigl[ x - \tau p(1) \Bigr]  & \tau\gg 1
  \end{cases}
\label{eq36}
\end{equation}
and
\begin{equation}
 p_{{\rm loss}}(t)
  =
  \begin{cases}
   p(1) \displaystyle\sqrt{\frac{4\tau}{\pi}} & \tau\ll1 \\[4mm]
   1  & \tau\gg 1
  \end{cases}
\label{eq36a}
\end{equation}
The conditional PDF (with the condition that the system dropped at least one packet during the time $t$) can be defined
as follows
\begin{equation*}
 w_{\rm loss}(x;t)
  \equiv\frac{p_{\rm loss}(x;t)}{p_{\rm loss}(t)}
  =
  \begin{cases}
   \displaystyle\sqrt{\frac{\pi}{4\tau}} {\rm erfc}\left[\displaystyle\frac{x}{\sqrt{4\tau}}\right] & \tau\ll1 \\[4mm]
   \delta\Bigl[ x - \tau p(1) \Bigr]  & \tau\gg 1
  \end{cases}
\end{equation*}

The central moments can be calculated in the same way as (\ref{eq13}). Here we will consider only the variance of the
losses $\sigma^2_{\rm loss}(t)$ in the limit $\tau\gg1$:
\begin{equation}
\begin{split}
 \sigma^2_{\rm loss}(t)
  &=
  m_{\rm loss}^{(1)}(t)
  \left[ \frac{1}{|v|}{\rm cotanh}|v| - \sinh^{-2}|v| \right]
\\&
  =
  \begin{cases}
   \displaystyle\frac{2}{3}m_{\rm loss}^{(1)}(t) & |v|\ll 1 \\[4mm]
   \displaystyle\frac{1}{|v|}m_{\rm loss}^{(1)}(t) & |v|\gg 1
  \end{cases}
\end{split}
\label{eq42}
\end{equation}
This is essentially in agreement with the result of considerations in Ref.~\onlinecite{Yur06} where a simple discrete-time model for studying losses in a single buffer was introduced. In that model packets of fixed size arrive with probability $p$ at the equidistant time epochs. The service was deterministic, and half of packet was served between the successive time epochs. In spite of such oversimplification, the discrete model has delivered quantitatively the same results which indicates the universality of the approach.

Finally, we calculate the correlator of the fluctuations of losses measured during two time intervals of length $t_1$
and $t_2$ correspondingly and separated by the time $T$:
\begin{equation*}
 {\rm corr}(t_1,t_2,T)
  = \int\limits_0^1\!{\rm d}\ell\ \rho(t_1,t_2,T)
   - m_{\rm loss}^{(1)}(t_1)m_{\rm loss}^{(1)}(t_2)
\end{equation*}
where
\begin{equation*}
\begin{split}
 &\rho(t_1,t_2,T)
\\
&\
 = r_{\rm loss}^2\int\limits_{0}^{t_1}\!{\rm d}t'_1\!\int\limits_{0}^{t_2}\!{\rm d}t'_2\
     w(1,t'_1 + t_2 - t'_2 + T;1) p(1)
\end{split}
\end{equation*}
with $r_{\rm loss}$ defined in (\ref{eq12}).

In the regime $T\gg t_1,t_2$ and $T\gg 2/\sigma^2$ it can be shown that
\begin{equation}
 {\rm corr}(t_1,t_2,T)
  \mathop{\rightarrow}\limits_{T\to\infty}0\ ,
\label{eq48}
\end{equation}
as we would expect. In fact, the correlator goes to zero exponentially if $v\neq0$. In the opposite regime
$2/\sigma^2\gg T\gg t_1,t_2$ we have
\begin{equation}
 {\rm corr}(t_1,t_2,T)
  = m_{\rm loss}^{(1)}(t_1)m_{\rm loss}^{(1)}(t_2)
  \frac{1}{p(1)}\sqrt{\frac{2}{\pi\sigma^2 T}}\,,
\label{eq49}
\end{equation}
which is again in agreement with the results of the discrete-time considerations \cite{Yur06}.

\section{Discussion and Conclusion}

As we would expect intuitively, loss events separated widely in
time are uncorrelated as shown by equation~(\ref{eq48}). By widely
separated in time, we mean that the time separation of the two
observation intervals in which losses occur is much longer than
the time over which fluctuations of queue length become comparable
or much bigger than the buffer size itself, i.e. $2/\sigma^2$.

However, in the case when the separation time is much smaller than
$2/\sigma^2$, the correlations of loss fluctuations are decaying
very, very slowly, as can be seen from equation~(\ref{eq49}). Such
time intervals are likely to be comparable or even smaller than
the round trip times for typical TCP connections. TCP is the
protocol that controls the rate at which data is sent across a
network, between a particular source and destination. The exact
details of the congeestion control operation of TCP can be found
in \cite{RFC2581}. For our purposes we shall only focus on its
salient congestion control features and the implications of the
result of equation~(\ref{eq49}) on it.

TCP limits its sending rate as a function of the perceived network
congestion. It operates on a virtual control loop of sending
packets, receiving acknowledgements and estimating the round trip
time. Once a packet is lost, the sender cuts its transmission rate
by half. If no other loss is detected it increases its sending
rate linearly by a small increment. But if a subsequent loss event
is detected it cuts its transmission rate in half again. If
successive loss events occur, which according to
equation~(\ref{eq49}) is likely on the relevant time scale, the
reduction in transmission rate can be dramatic and potentially
unnecessary. As there are multiple TCP connections experiencing losses at
the same buffer this will lead to a cycle of rapid under-usage and
slow convergence to congestion, which is clearly undesirable and
ineffective.

Studying of spatial correlations of loss fluctuations over a
network is work in progress. This will help us quantify the second
significant aspect of TCP operation which is its reaction to
time-out events, as this is connected to correlated losses and
delays around the sequence of buffers forming each control loop.

To conclude, we emphasize that the stability of a network with respect to data loss was mostly analyzed in the past from the viewpoint of the loss of physical connectivity in the network topology where a failure of a given node or link was treated as a (probabilistic) input into a network model. Here we have studied \textit{dynamical} fluctuations in data loss in a single node (memory buffer) of the network. We have shown that the strong fluctuations and long-time memory in losses inevitably follow from the discrete character of signal propagation in the packet-switched networks. This single-node fluctuations can potentially trigger a cascade of failures in neighboring nodes and thus result in a temporal failure of large parts of the network. In the next stage, we intend to utilize these features of the local data loss as dynamical inputs into the network and thus study possible abrupt increase of data loss in the network triggered by a local overload.

\acknowledgements{This work is supported by EPSRC grant GR/T23725/01.}

\end{document}